\documentclass[runningheads]{llncs}
\usepackage{graphicx}
\usepackage{listings}
\usepackage{array}
\usepackage[hidelinks]{hyperref}
\begin{document}
\title{Role of Apache Software Foundation in Big Data Projects}
%
%
\author{Aleem Akhtar}

	\institute{Virtual University of Pakistan \\
		\email{aleem.akhtar@seecs.edu.pk} \\
		 }
%
%
%
\maketitle              

\begin{abstract}
	With the increase in amount of Big Data being generated each year, tools and technologies developed and used for the purpose of storing, processing and analyzing Big Data has also improved. Open-Source software has been an important factor in the success and innovation in the field of Big Data while Apache Software Foundation (ASF) has played a crucial role in this success and innovation by providing a number of state-of-the-art projects, free and open to the public. ASF has classified its project in different categories. In this report, projects listed under Big Data category are deeply analyzed and discussed with reference to one-of-the seven sub-categories defined. Our investigation has shown that many of the Apache Big Data projects are autonomous but some are built based on other Apache projects and some work in conjunction with other projects to improve and ease development in Big Data space.
\end{abstract}

\section{Introduction}
Last decade has seen an explosion of Data. Huge amount of data is being produced at very hight rate from Internet sites, Government records, scientific experiments, sensor networks, and many other sources like online transactions, images, audio, videos, posts, health records, emails, logs, click streams, social networks, mobile phones, and their apps ~\cite{dean2008mapreduce}\cite{schneider2012custom}. Such data cannot be managed or processed in reasonable amount of time by traditional set of database tools, therefore, Big Data term was introduced for such data. Until 2005, 5 Exabyte of Data was generated but now 2.5 quintillion bytes of data is produced in a single day \cite{marr2018}. 2.72 zeta bytes of data was generated by digital world till 2012, and after doubling every year it reached to 8 zeta byte in 2015 \cite{garlasu2013big} and by the end of 2020 it is expected to reach 44 zettabytes, or 44 trillion gigabytes \cite{turner2014digital}. As per SINTEF report in 2013, 90\% of this data was produced in just two years \cite{petter2013}\cite{seo2010hama}. Genome decryption process used to take nearly 10 years in past, now is done in less than a week \cite{bhardwaj2016survey}. Multimedia data increased by 7\% by 2013 \cite{manyika2011big}. With servers in millions, Google is largest Internet Company. More than 10 billion text messages are sent by 7 billion mobile subscribers’ every day. Movies and the sharing platforms are expected to have nearly 50 billion movies connected by the end of this year. 

This amount of information expected to increase by 50 times in next decade with technology experts to keep up with all data are expected to increase by 1.5 times \cite{tankard2012big}. This huge amount of data is increasing on daily basis with no end in sights. The need to store, process, and analyse this data is stronger than ever. Many tools are specifically built to store, process and analyse this data are being developed.

\subsection{Big Data}
\begin{figure}[htbp]
	\centerline{\includegraphics [width=0.47\textwidth]{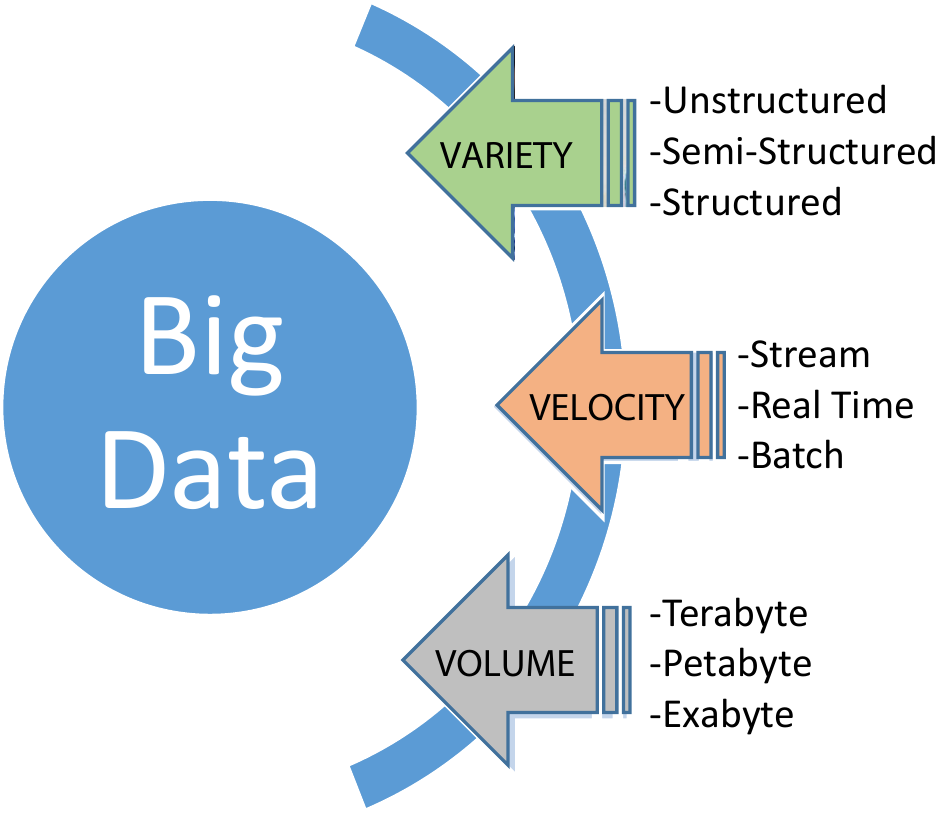}}
	\caption{Big Data Three V's}
	\label{threev}
\end{figure}

A revolutionary step required by Big Data from traditional analytics, defined three main components called three V’s of Big Data: variety, velocity and volume as shown in fig \ref{threev} \cite{gerhardt2012unlocking}\cite{center2012planning}\cite{schneider2012custom}\cite{seo2010hama}.

\begin{itemize}
	\item \textbf{Volume:} it defines size of data generally larger than terabytes. Traditional store and analysis techniques are outstripped by this grand scale of data \cite{dean2008mapreduce}\cite{madden2012databases}. 
	\item \textbf{Variety:} It defines how data elements are related to each other. In structured data, tags are present so data elements can easily be separated, whereas in unstructured data, due to randomness, it is very difficult to analyze. Semi-structured data does not have fixed fields but can easily be separated \cite{dean2008mapreduce}\cite{seo2010hama}. 
	
	\item \textbf{Velocity:} It defines speed of data at which it is being generated. It can be real-time, streamed or in batches \cite{dean2008mapreduce}\cite{madden2012databases}. 
\end{itemize}

In many literature, there is fourth component ‘Verification’ also discussed. Due to intensity of information security feature is required as controlling large data is not easy.

\subsection{Open Source Software}
Many open source software in the Big Data field is very crucial and many Big Data projects are being made open and free to the general public. Major dominating industry in Big Data solutions is open-source software and giants like IBM, Oracle and Microsoft are now following the footsteps to make their proprietary software as open-source software. There is rapid change in innovation of Big Data field and solutions due to development in open-source software.

Richard Stallman started open-source movement in 1983 with development of GNU project \cite{stallman1986free}. Information science research has well portion of open-source software and open-source communities are using development methods which are proven quite successful. A very important factor for success of any open-source project is community around that project which mainly gives development and innovation of project software solutions that are diverse and robust are generally supported by well-functioning diverse communities.

\section{Apache Software Foundation}
The Apache Software Foundation's history is connected to the Apache HTTP Server, which began in February 1993. A team of eight developers –later called as Apache Group– started working to expand the NCSA HTTPd Daemon. The Apache Software Foundation was established on March 25, 1999 \cite{ASFCertificate}. On April 13, 1999, the Apache software Foundation’s first official meeting was held. The early members of the Apache Software Foundation were: Miguel Gonzales, Ken Coar, Brian Behlendorf, Mark Cox, Ralf S. Engelschall, Paul Sutton, Marc Slemko, Lars Eilebrecht, Dean Gaudet, Sameer Parekh, Roy T. Fielding, Cliff Skolnick, Jim Jagielski, Ben Hyde, Alexei Kosut, Martin Kraemer, Doug MacEachern, Ben Laurie, , Aram Mirzadeh, William (Bill) Stoddard, Dirk-Willem van Gulik, and Randy Terbush \cite{ASF1999Meeting}. Board members were elected after a series of other meetings. After dealing with other legal issues related to the formation of the company, June 1, 1999 was set as the effective date of the Apache Software Foundation \cite{ASFMeetingJune99}.

Software development activities at Apache are divided into semi-autonomous areas known as “top-level projects” with some of them comprised of sub-projects (previously called “Project management committee in the bylaws \cite{ASFBylaws}).   Dissimilar to other free-and-open-source projects hosting organizations, project is to be licensed to ASF before it is hosted at Apache with contributor or grant agreements \cite{ASFLicense}. As such, ASF acquire the right of intellectual property, needed to develop and distribute all of its projects \cite{amant2008handbook}.

\subsection{Powered By Apache}
Every Internet-connected country of the world is using Apache software. ASF projects serve as the backbone for some of most widely used and visible applications of the world such as Big Data, Deep Learning \& Artificial Intelligence, Cloud Computing, DevOps, build management, IoT and Edge computing, content management, servers, mobile and web frameworks, and among many other similar fields \cite{ASFPoweredBy}. List of applications that are "Powered by Apache" include:

\begin{itemize}
	\item	\textbf{NASA:} powering Ocean Science and Big Earth  data analytics;
	\item	\textbf{NASA Jet Propulsion Laboratory:} accessing content across multi-mission, multi-instrument science data systems;
	\item \textbf{Panama Papers:} document, search and library management tools used in the nearly 3TB Pulitzer Prize-winning investigation;
	\item\textbf{IBM Watson:} advancing semantics capabilities and data intelligence to win first-ever "Man vs. Machine" competition on Jeopardy!
	\item	\textbf{Facebook:} requests processing at 300PB data warehouse, connecting more than 2 billion active users;
	\item	\textbf{Twitter:} processing and analyzing of more than 200B annual tweets in Zettabytes;
	\item	\textbf{Adobe:} powering core of Experience Manager and I/O Runtime;
	\item	\textbf{Netflix:} data ingestion pipeline and stream processing 3 trillion events each day;
	\item	\textbf{Minecraft:} libraries bundling for modification of the all time second most popular video game;
	\item	\textbf{Amazon Music:} 16M+ subscribers and tuning recommendations; 
	\item	\textbf{AOL}: ingesting more than 20 Terabyte of daily data;
	\item	\textbf{Formula 1, Daimler, and Audi:} real time data streaming in vehicles; 
	\item	\textbf{Pinterest:} processing more than 800 billion daily events;
	\item	\textbf{Uber:} handling 1M writes per second for 99.99\% availability to users and drivers;
	\item \textbf{Mobile app developers:} unifying mobile application development across iOS, Android, Windows Mobile and Blackberry operating systems;
	\item	\textbf{European Space Agency:} powering next-generation simulators infrastructure and new mission control system;
	\item \textbf{US Federal Aviation Administration:} system-wide information management to enable every airplane take off and land in US airspace;
	
\end{itemize}

\section{ASF Big Data Projects}
Apache Software Foundation list down projects in three main categories: active projects are those which are currently available for downloads and constantly being updated. Projects which are no longer provided with Apache support but are still governed by Apache license are sent in to the attic and are retired. Finally, Apache Incubator service provides an entry path for codebases and projects desiring to become part of the Apache Software Foundation \cite{ASFProjectsList}. Currently there are 50 projects listed under Big Data category on Apache website with 44 of them being active, 03 retired and 03 currently in incubation stages \cite{ASFBigDataProjects}. Table \ref{tab1} presents a list of Big Data projects.

\begin{figure*}[htbp]
	\centerline{\includegraphics [width=0.97\textwidth]{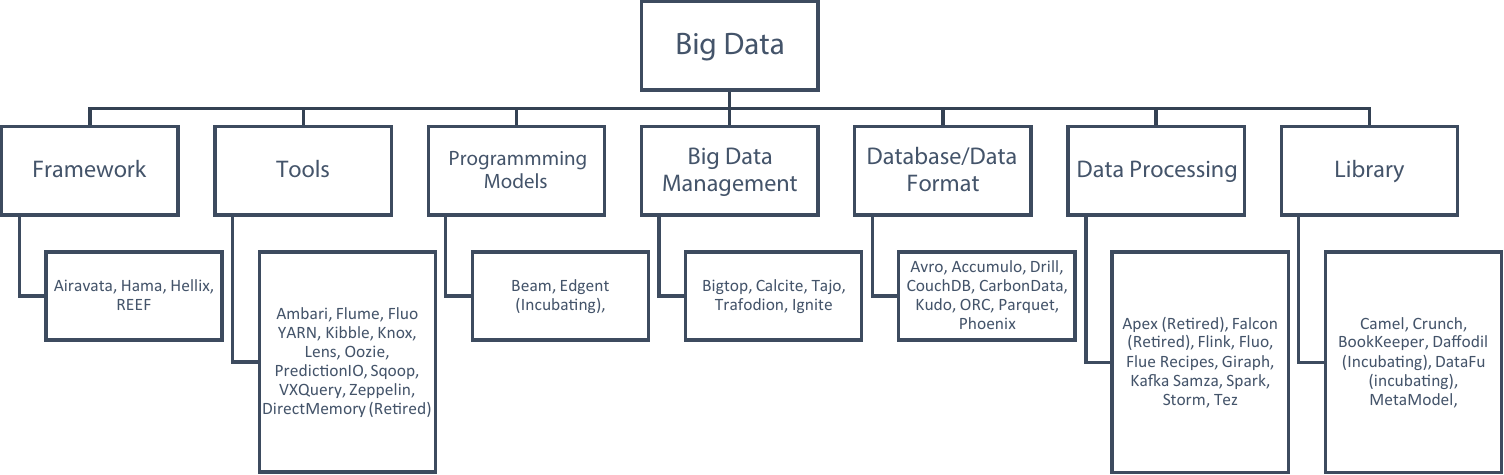}}
	\caption{Sub-Categories of Apache Big Data Projects}
	\label{bigdataprojects}
\end{figure*}

\begin{table}[htbp]
	\caption{List of ASF Big Data Projects}
	\begin{center}
		\begin{tabular} {|p{0.20\linewidth}|m{0.80\linewidth}|}
			\hline
			\centering\textbf{Active} & {Accumulo, Airavata, Ambari,Avro, Beam, 
				Bigtop, BookKeeper, Calcite, Camel, CarbonData, CouchDB, 
				Crunch, Drill, Flink, Flume, Fluo, Fluo Recipes, Fluo YARN,
				Giraph, Hama, Helix, Ignite, Kafka, Kibble, Knox, Kudu,
				Lens, MetaModel, OODT, Oozie, ORC, Parquet, Phoenix, 
				PredictionIO, REEF, Samza, Spark, Sqoop, Storm, Tajo, 
				Tez, Trafodion, VXQuery, Zeppelin } \\
			
			\hline
			\centering\textbf{Retired} & Apex, DirectMemory, Falcon \\
			
			\hline
			\centering\textbf{Incubation} & {Daffodil, DataFu,  Edgent} \\
			\hline
		\end{tabular}
		\label{tab1}
	\end{center}
\end{table}
These projects are further divided into sub-categories based on services they provides. Analyzing each project in detail, a sub-category list is prepared in fig-\ref{bigdataprojects}, followed by brief explanation of each sub-category.

\subsection{Frameworks}
Apache Software Foundation is full of open-source projects that serve as Frameworks to efficiently manage resources, jobs, workflows, applications running on clusters. Apache Airavata \cite{marru2011apache} is a micro-service architecture based software framework for managing and executing computational workflows and jobs on distributed computing resources including commercial clouds, local clusters, national grids, supercomputers and academic clouds. Prevailing use of Airavata is to build web-based science gateways and assisting in composing, monitoring, execution and management of large-scale applications and workflows wrapped or composed of web-services.

The Apache Hama \cite{siddique2016apache} is a scalable and efficient general-purpose Bulk Synchronous Parallel (BSP) computing engine which is used in Big Data Analytics with a purpose to speed-up diverse set of compute intensive analytics applications. The Apache Helix \cite{ApacheHelix} is a general-purpose cluster management framework which is used to automatically manage resources that are replicated, partitioned and distributed on cluster of nodes. In case of cluster expansion, node failure and recover, or cluster reconfiguration, Helix automatically reassign resources. Apache Retainable Evaluator Execution Framework or simply Apache REEF \cite{chun2017apache} is a development framework that provides mechanism to simplify the development of Big Data applications on cloud platforms supporting Resource Manager Service like Apache Hadoop YARN or Apache Mesos.

\subsection{Tools}
There are variety of tools developed around Big Data projects to facilitate in processing and management of large amounts of data. The Apache Ambari initially started as sub-project of Hadoop to provide system administrators facilities of provision, monitoring and management of Hadoop clusters. Apache Ambari \cite{wadkar2014apache} is now a top-level project managed by its own community to facilitate integration of Hadoop with prevailing enterprise infrastructure. The Apache Flume \cite{hoffman2013apache} and Apache Sqoop \cite{vohra2016using} are two reliable, distributed and available systems to efficiently collect, move, and aggregate and huge chunks of log data from number of various sources to a centralized data store.

The Apache Fluo YARN \cite{ApacheFluoYarn} is a sub-project of Apache Fluo in the form of a tool to run Apache Fluo applications in Apache Hadoop YARN. The Apache Kibble \cite{ApacheKibble} is a tools suite similar to Apache Flume to collect, visualise, and aggregate software projects activities, Apache Zeppelin \cite{cheng2018building} is a web-based tool for data scientists with similar features. To have Hadoop clusters a REST interactions Apache Knox \cite{knox2019rest} provide a REST API Gateway that provides important features helping to control, monitor, integrate and automate enterprise’s critical analytical and administrative requirements. Apache Lens \cite{koitzsch2017relational} integrates traditional data warehouses and Hadoop to provide a single data view across optimal execution environment and multi-tiered data stores. To schedule Apache Hadoop jobs like Java Map-reduce, Hive, and Pig along with system specific jobs like shell scripts and Java programs, Apache Oozie \cite{islam2015apache} is integrated with Hadoop stack.

Apache PredictionIO \cite{ApachePredictionIO} and Apache VXQuery \cite{carman2015apache} are two very useful tools with one providing machine learning services and other being XML Query processor, respectively. PredictionIO let developers to deploy and manage production-ready predictive services for machine learning jobs. VXQuery uses cluster to evaluate queries on huge sets of comparatively small XML documents.

\subsection{Programming Model}
Apache Software Foundations provides a handful set of very useful programming models and runtime that provide facility of running data processing jobs on distributed and diverse execution engines. To run both stream and batch data processing, Apache Beam \cite{beam2017apache} provides a unified programming model. Users develop application programs in the form of pipelines using Apache Beam SDKs and Beam’s supported processing back-ends like Apache Spark, Apache Flink, and Apache Apex execute those pipelines. Apache Edgent \cite{edgent2017v1} is an incubating project and still not a top-level project but concept behind this project is to provide a micro-kernel style runtime and programming model that can easily be embedded in edge devices to enable real-time, local, analytics on continuous stream of data originating from vehicles, equipment, appliances, systems, sensors and devices of all kinds like smart phones and Raspberry Pi. 

\subsection{Big Data Management}
Big Data Management is an important aspect in Big Data field and Apache Software Foundation hosts open-source projects that provide comprehensive set of big data management features either by integrating on top of Hadoop cluster or just by simply providing query transformation rules. Apache Bigtop \cite{ApacheBigtop} and Apache Trafodion \cite{ApacheTrafodion} are two main big data management projects that provide features of development of applications to run on Hadoop ecosystem. Trafodion extends Hadoop to provide transactional integrity for new applications to run on Hadoop whereas Bigtop let packaging and testing of Hadoop-related projects. Apache Phoenix \cite{akhtar2016pro} is another project that can easily integrate in the Hadoop ecosystem and other Apache products like Flume, Spark, Map Reduce and Pig.

Apache Calcite \cite{begoli2018apache} and Apache Tajo \cite{tajo2013big} are somehow similar in providing frameworks to process web-scale data sets. Calcite uses transformation rules to convert relational algebra queries into efficient executable form with no initial cost models. While primary goal of Tajo is to use progressive query and cost-based optimization techniques to provide dynamic load-balancing and fault-tolerance to run long-queries. Apache Ignite \cite{ApacheIgnite} uses caching platform and In-Memory Database to deliver high performance. 

\subsection{Libraries}
List of library projects providing different kinds of services at ASF is very long and libraries specifically designed for Big Data projects is also quite comprehensive, however, only those libraries which are widely used discussed here. Apache BookKeeper \cite{ApacheBookKeeper} is a highly available and scalable replicated log service that can turn any individual service into replicated service. Based on Enterprise Integration patterns, Apache Camel is another great integration library.

To parse fixed data, Apache Daffodil uses DFDL data specifications and output fixed format data into infoset in the form of JSON or XML. Other technologies working on JSON or XML can easily utilize this output. Daffodil can also serialize or reverse-parse JSON or XML infoset into fixed data format. Apache MetaModel \cite{ApacheMetaModel} provides a query API and uniform connector to many datastore types, including: JSON files, XML files, CSV files, fixed with files, Excel spreadsheets, Apache Cassandra, Apache HBase, Apache CouchDB, MongoDB, Relational (JDBC) databases, SugarCRM, Plain Old Java Objects (POJO) and ElasticSearch. Computations happening at regular intervals generally have unnecessarily repetitions, therefore, Apache DatFu \cite{ApacheDataFu} reduces up to 95\% of computational resources by making computations more efficient.

\subsection{Database/Data Format}
For fast processing of data, efficient database and data storage format are very important and ASF outlines a comprehensive list of such projects. There is a complete range of Apache databases for different use cases. However, Apache CouchDB \cite{couchdb2010apache} is one such project that can efficiently work with both web and mobile apps. Effective distribution of data using incremental replication is one of the Apache CouchDB’s main feature. Apache Accumulo \cite{ApacheAccumulo} and Apache Drill \cite{ApacheDrill} are two key projects connected with Google products with Accumulo based on Google’s BigTable design and Drill partially based on Google’s Dremel. Accumulo was built on top of Apache Hadoop, Apache Thrift and Apache Zookeeper with BigTable design improvements. A variety of NoSQL databases and file systems are supported by Drill, including HDFS, MapR-DB, MapR-FS, MongoDB, HBase, Azure Blob Storage, Amazon S3, NAS, Swift, Google Cloud Storage and local files. Unlike Accumulo, Apache Avro \cite{ApacheAvro} is developed within Hadoop to provide data serialization and row-oriented remote procedure call framework. Data types and protocols are defined using JSON and serialized in compact binary format.

Apache CarbonData \cite{ApacheCarbonData}, Apache Kudo \cite{ApacheKudo}, Apache ORC \cite{ApacheOrc} and Apache Perquet \cite{ApacheParquet} are some of the open-source projects that work with columnar storage file format. CarbonData uses advanced index, encoding and compression techniques to improve computing efficiency. To support Apache Hadoop platform, Kudo is developed as columnar storage manner whereas to efficiently optimize large streaming workloads at Hadoop, ORC is designed as type-aware columnar file format. Apache Parquet another Hadoop supported generic columnar storage format which can be used with any data model, processing framework, or programming language.  

\subsection{Data Processing}
Big Data Analytics is one of the most important objective sought after by academic institutions, researchers, scientists, and companies. For this purpose, data processing projects developed and maintained by Apache are frontrunners and are used by many large enterprises. As specified in introduction section, one component of Big Data is velocity, and data can be produced in batch, stream or run-time, therefore, data processing framework must be able to efficiently process it for best results. For this purpose, some Apache projects process data only in batches, some in streams, and some in both formats. Apache Flink \cite{carbone2015apache} works with data in large batches and it combines the programing flexibility and scalability of distributed MapReduce-like platforms with the query optimization, out-of-core execution, and efficiency capabilities found in parallel databases. Apache Fluo \cite{ApacheFluo} is another distributed batch processing system built on Apache Accumulo. With Fluo, new data can easily be joined with large existing data without reprocessing of entire data. Apache Fluo Recipe is built on Apache Fluo but with additional features and is maintained separately with independent releases. 

Apache Spark \cite{zaharia2016apache}, Apache Samza \cite{ApacheSamza}, Apache Storm \cite{ApacheStorm} and Apache Kafka \cite{ApacheKafka} are all open-source stream processing platforms with Spark providing features of batch processing as well. High Level APIs are provided by Spark in Scala, Java, R and Python for fast data processing along with libraries for graph analytics, machine learning and stream processing. Kafka developed by LinkedIn and donated to ASF provides low-latency, high throughput, unified platform for handling real-time feeds. Kafka stream data is used by Samza for processing with the use of messages. Apache Storm provides general primitives for processing real-time data.

Apache Giraph \cite{ApacheGiraph} and Apache Tez \cite{ApacheTez} are two graph processing systems used for data processing. Social graph formed by users at Facebook are analyzed using Giraph. Whereas, to process complex directed-acyclic-graphs (DAGs) of data-processing, Tez is widely used.

\section{Discussion}
Apache Software Foundation (ASF) \cite{ASFMain} is a non-profit open-source software foundation, which is considered a very important organisation in Big Data space. There is a diverse software development at the foundation, and many widely used software projects are placed at it, with user community expanded to all over the world. Key reason behind success of ASF is importance of community above other requirements. Because of this, an agile and flexible environment is provided at ASF for development of open-source projects. To maintain successful projects, legal framework and infrastructure is also provided. In last few years, a large chunk of successful Big Data projects have been attracted towards ASF. 

Github \cite{Github} is another leading open-source software platform for Big Data projects. Unlike ASF, Github is a git-based code repository and does not provide organisational and legal framework. Github is an ad-hoc platform and for successful projects, communities are formed around projects in ad-hoc manner. Github is used for hosting open-source projects by Universities, foundations like ASF, and companies like Netflix and LinkedIn. LinkedIn \cite{LinkedIn} and Netflix \cite{Netflix} are first companies to make their code open-source to public. Large software companies like Facebook, Yahoo and Twitter create projects and donate to public through open-source software foundations. Both original software developer and community get benefitted by this process. Products evolve very quickly and mature fast when software creators expose their code to diverse communities. Product become resilient by being battle tested in all kinds of scenarios for free. Trust and high credibility is created among peer developers for the leaders of open-source software projects is one of the most rewarding thing about making software open-source.

Apache Software Foundation is an all-volunteer community with more than 700 individual members and nearly 7,000 committers working on 200+ million lines of code in more than 350 open source projects that are used by billions of the users and developers across the globe. There are more than 350 projects that ASF has provided free to the public. Nearly 30 million page visits per week by developers and users are recorded at ASF’s official website and its sub-domains. Excluding convenient binaries, source code from Apache Mirrors have been downloaded about 9M+ times. 

\begin{figure*}[htbp]
	\centering \includegraphics [width=0.97\textwidth]{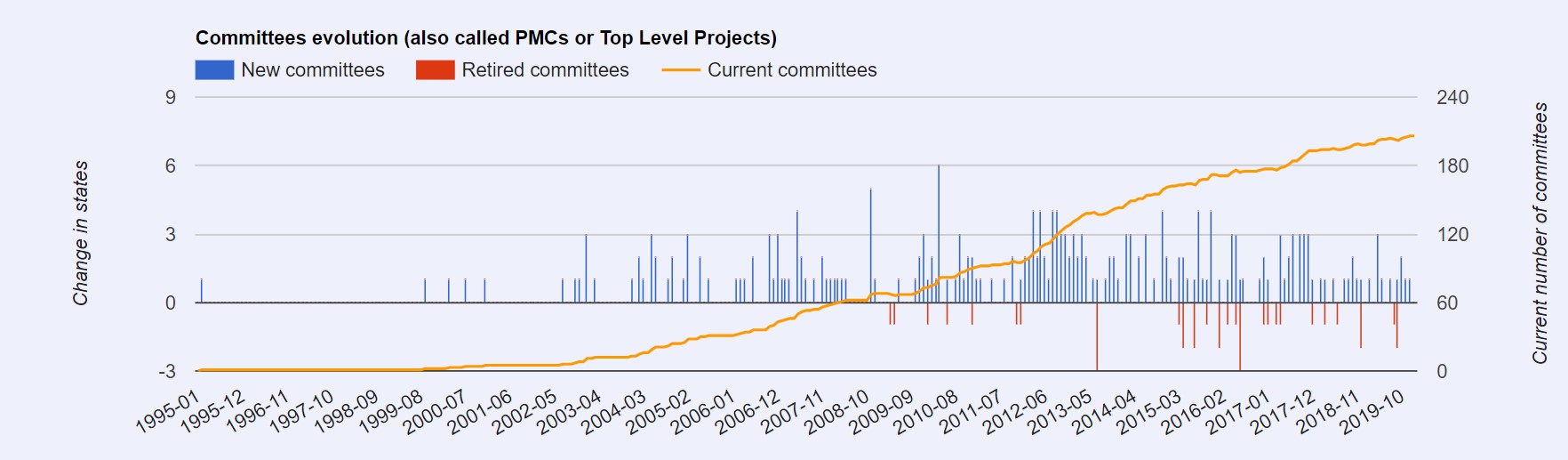}
	\caption{Timeline of Apache Committees Evolution}
	\label{committees}
	
	\centering	\includegraphics  [width=0.97\textwidth] {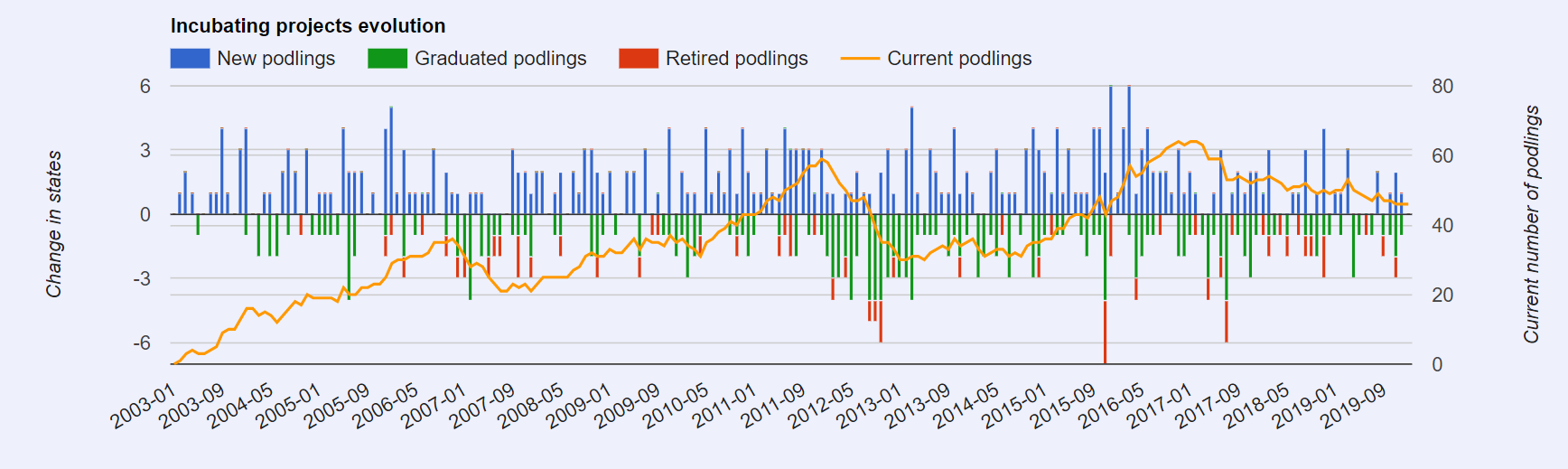}
	\caption{Timeline of Apache Incubating Projects Evolution}
	\label{incubating}
\end{figure*}

First Big Data project launched in January, 2008 was Apache Hadoop while first Big Data project to retire was Apache DirectMemory in July, 2015. Fig \ref{committees} presents a timeline graph of Apache Committees evolution with Apache HTTP Server being first committee to launch in 1995 while Apache Druid being latest committee to launch in last month of 2019. Fig \ref{incubating} presents an evolution of apache incubating projects. 

\begin{figure}[htbp]
	\centerline{\includegraphics [width=0.47\textwidth]{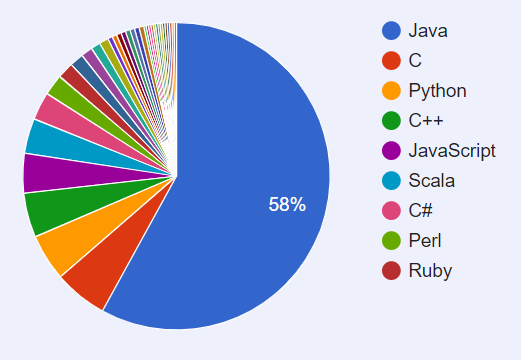}}
	\caption{Apache Projects Language Distribution}
	\label{languages}
	\centerline{\includegraphics [width=0.47\textwidth]{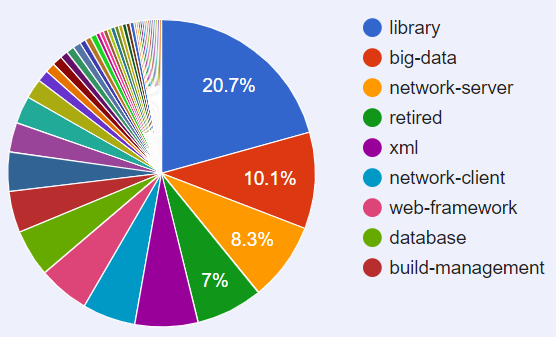}}
	\caption{Projects Categories}
	\label{categories}
\end{figure}

Fig \ref{languages} presents language distribution of Apache projects. Java being the major language for most of the projects with nearly 58\% projects being developed in Java. There are many projects which are provided in more than one language like Apache Spark which is available in Java, Scala and Python. Fig \ref{categories} gives a brief project categories distribution with nearly 21\% of the Apache projects being libraries followed by Big Data \cite{ASFMain}.

\section{Conclusion, Related Work, and Future Work}
\subsection{Conclusion}
In this report, we deeply investigated Apache Software Foundation for open-source projects which are listed under Big Data category. Each project was studied to know its sub-category. Seven sub-categories were identified and projects were investigated to understand relationship among each of them. Frameworks, Tools, Programming Models, Big Data Management, Libraries, Database/Data Format, and Data Processing are main sub-categories studied and presented in this report. Our investigation showed that, many projects work independently while there are some of the projects that either utilize services of other Apache projects or provide services to some other projects for better performance and ease of use. 

Some of the projects discussed in this report, have support of large competing technology organizations. Even with this fact, projects are using and complimenting each other and co-exist to provide an exceptional open development environment in the big data space for advanced and state-of-the-art projects. Many successful and important open projects are now permanent member of Apache, and newer projects are attracted towards Apache with an increasing pace.

\subsection{Related Work}
There is a lot of research done on individual Apache projects especially on data processing frameworks identifying performance, use-cases, potential issues, and future targets. There is also research available that compare and highlight performance between two or more similar Apache projects. However, there is not much research done on complete Big Data projects list. Kamburugamuve provided a similar research report as part of PhD qualifying exam in which he presented Apache Big Data projects in the form of a layered architecture \cite{kamburugamuve2013survey}. This report was presented in 2013, and many new big data projects were launched after that. In this report, we covered a bigger set of projects.
\subsection{Future Work}
Principal Analyst at RedMonk Stephen O’Grady appreciated ASF by saying that “The Apache Software Foundation has been one of the few institutions that have been crucial for growth and advancement of Open Source projects in the last two decades. A neutral environment is provided to developers with different backgrounds to work together, which has mainly played a very important role in open source success and ASF is looks determined to continue to play similar role in next decade.” In this report, we only covered projects that are listed under Big Data tag on Apache official project website. Future work, will include more detailed analyses of other project categories that overlap somehow with Big Data field but are not mentioned in Big Data category. Another future research perspective is to deeply investigate data processing models/frameworks and provide a comprehensive comparison based on performance, ease-of-use, and other key metrics.

\bibliographystyle{IEEEtran}


\bibliography{ASFinBigDataProjects}

\end{document}